\documentclass[useAMS, usenatbib, onecolumn]{mn2e}
\usepackage{epsfig}
\def\be{\begin{equation}}
\def\ee{\end{equation}}
\def\ba{\begin{eqnarray}}
\def\ea{\end{eqnarray}}
\def\go{\mathrel{\raise.3ex\hbox{$>$}\mkern-14mu
             \lower0.6ex\hbox{$\sim$}}}
\def\lo{\mathrel{\raise.3ex\hbox{$<$}\mkern-14mu
             \lower0.6ex\hbox{$\sim$}}}

\def\tomega{{\tilde{\omega}}}

\begin{document}
\title[Corotational Damping of C-Modes]
{Corotational Damping of Diskoseismic C-modes in Black Hole Accretion Discs}
\author[D. Tsang and D. Lai]
{David Tsang$^{1,2}$\thanks{Email:
dtsang@astro.cornell.edu; dong@astro.cornell.edu} 
and Dong Lai$^{1}$\footnotemark[1] \\ 
$^1$Department of Astronomy, Cornell University, Ithaca, NY 14853, USA \\
$^2$Department of Physics, Cornell University, Ithaca, NY 14853, USA \\}

\label{firstpage}
\maketitle

\begin{abstract}
Diskoseismic c-modes in accretion discs have been
invoked to explain low-frequency variabilities observed
in black-hole X-ray binaries. These modes are 
trapped in the inner-most region of the disc and
have frequencies much lower than the rotation frequency
at the disc inner radius. We show that because
the trapped waves can tunnel through the evanescent barrier to the corotational
wave zone, the c-modes are damped due to wave
absorption at the corotation resonance. We calculate the corotational
damping rates of various c-modes using the WKB approximation.
The damping rate varies widely depending on the mode frequency, the black hole
spin parameter and the disc sound speed, and is generally much less than $10\%$ of
the mode frequency. A sufficiently strong excitation mechanism is needed
to overcome this corotational damping and make the mode observable.
\end{abstract}

\begin{keywords}
accretion, accretion discs -- hydrodynamics -- waves -- 
-- black hole physics -- X-rays: binaries
\end{keywords}
\section{Introduction}

Diskoseismic oscillations of accretion discs around relativistic
objects have been studied for over two decades (e.g., Kato \& Fukue
1980; Okazaki et al. 1987; Nowak \& Wagoner 1991, see Wagoner 1999,
Kato 2001 for reviews), and have been used as models for the time
variability and quasi-periodic oscillations (QPOs) in X-ray emissions
from black-hole X-ray binaries.

The c-modes (or so-called corrugation waves) were first proposed to
explain low-frequency variabilities as their oscillation frequencies
are lower than the associated g-modes and p-modes (see section
3). Kato (1983) and Okazaki \& Kato (1985) showed the existence of
one-armed ($m=1$), low-frequency modes in nearly Keplerian (Newtonian) discs,
while later work (Kato 1989; Silbergleit et al.~2001) demonstrated the
presence of low-frequency c-modes in relativistic accretion discs,
particularly the one-armed corrugation waves with a single node 
($n=1$; see section 3 below) in the
vertical direction, which oscillate at (approximately) the
Lense-Thirring precession frequency evaluated at the outer edge of the
trapping region.

In a recent paper (Lai \& Tsang 2008), we studied the global
corotational instability of non-axisymmetric p-modes ($n=0$) in black hole
accretion discs.  The mode is trapped inside the corotation resonance
radius $r_c$ (where the wave pattern rotation speed $\omega/m$ equals
the disc rotation rate $\Omega$) and carries a negative energy.  We
showed that when the mode frequency $\omega$ is sufficiently high,
positive wave energy is absorbed at the corotation resonance, leading
to the growth of mode amplitude. The mode growth is further enhanced
by wave transmission beyond the corotation barrier.
Non-axisymmetric g-modes, on the other hand, may contain 
a corotation resonance in the wave zone. 
Kato (2003) and Li, Goodman \& Narayan (2003) showed that
such g-modes are heavily damped as the wave propagate through the corotation
resonance (see also Zhang \& Lai 2006). 

Diskoseismic c-modes are trapped in the inner most regions of black
hole accretion discs. Although their primary wave zones are 
separated from the corotation resonance, the wave can tunnel through 
the evanescent barrier and propagate again around the corotation. 
In this paper we calculate the analytic damping rate of c-modes due to
wave absorption at the corotation resonance. In section 2 we briefly 
review the basic properties of perturbations in a thin isothermal disc, and
present the linear perturbation equations. In section 3 we
discuss the propagation regions associated with various
diskoseismic modes, while in section 4 we demonstrate the effect of
the corotation resonance on wave propagation.  In section 5 the effect of the
corotation on the c-mode is studied and the
c-mode damping rates are calculated for different disc parameters. 
Section 6 contains our conclusion.

\section{Basic Setup and Equations}

Consider a thin isothermal disc with the unperturbed velocity
${\bf u}_o = (0, r\Omega, 0)$ in the cylindrical coordinates.
The vertical density profile is given by (for small $z \ll r$) 
\be 
\rho_o(r,z) = \frac{\Sigma(r)}{\sqrt{2\pi} H} \exp\left(-z^2/2H^2\right).
\ee 
Here $\Sigma(r)$ is the (vertically integrated) surface density and
$H=c_s/\Omega_\perp$ is the vertical scale height, where
$\Omega_\perp$ is the vertical oscillation frequency and $c_s$ is the
isothermal sound.

Perturbing the mass and momentum conservation equations gives
\ba
\frac{\partial}{\partial t}\delta \rho + \nabla\cdot(\rho_o \delta
{\bf u} + {\bf u}_o \delta \rho) &=& 0~,\\
\frac{\partial}{\partial t}\delta {\bf u} + ({\bf u}_o \cdot \nabla)
\delta {\bf u} + (\delta {\bf u} \cdot \nabla) {\bf u}_o &=& -\nabla
\delta h~,
\ea
with the enthalpy perturbation $\delta h \equiv \delta P/\rho = c_s^2
\delta \rho/\rho$, where we assume that the perturbations are also
isothermal.
Assuming perturbations of the form $\delta P, \delta {\bf u}, \delta
\rho \propto \exp(im\phi - i\omega t)$, we have
\ba
-i\tomega \frac{\rho_o}{c_s^2}\delta h +
\frac{1}{r}\frac{\partial}{\partial r}(r\rho_o \delta u_r) +
\frac{im}{r}\rho_o \delta u_\phi + \frac{\partial}{\partial z}(\rho
\delta u_z) &=& 0 \label{fullmomentumpert}\\
-i\tomega \delta u_r - 2\Omega \delta u_\phi &=& -\frac{\partial}{\partial r} \delta h
\label{eq:dur}\\
-i\tomega \delta u_\phi + \frac{\kappa^2}{2\Omega} \delta u_r &=& -\frac{im}{r}\delta h
\label{eq:duphi}\\
-i\tomega \delta u_z &=& -\frac{\partial}{\partial z}\delta h\label{zpert}
\ea
where $\tomega = \omega - m\Omega$.
Following Okazaki et al.~(1987), we assume a $z$-dependence of the
perturbations such that $\delta h, \delta u_r, \delta u_\phi \propto
{\rm H}_n(z/H)$ where ${\rm H}_n(z/H)$ is the Hermite polynomial
of order $n$.
Then equation (\ref{fullmomentumpert}) reduces to
\ba
-i\tomega \frac{\rho_o}{c_s^2}\delta h + \frac{1}{r}\frac{\partial}{\partial r}(r\rho_o \delta u_r) + \frac{im}{r}\rho_o \delta u_\phi -  \frac{n\rho_o}{i\tomega} \delta h= 0
\label{eq:dh}
\ea
Neglecting terms proportional to $dH/dr \sim O(1/r)$, and eliminating
the velocity perturbations $\delta u_r$ and $\delta u_\phi$ from equations
(\ref{eq:dur}), (\ref{eq:duphi}) and (\ref{eq:dh}), we obtain 
(see eq.~[29] in Zhang \& Lai [2006])
\be
\frac{d^2}{dr^2}\delta h - \left(\frac{d}{dr}\ln \frac{D}{r\Sigma}
\right)\frac{d}{dr}\delta h +
\left[\frac{2m\Omega}{r\tomega}\frac{d}{dr}\ln \frac{D}{\Omega\Sigma}
- \frac{m^2}{r^2} - \frac{D(\tomega^2 - n\Omega_\perp^2)}{c_s^2
\tomega^2}\right] \delta h = 0~, \label{mastereq} 
\ee 
where $D = \kappa^2 - \tomega^2$. This is our basic working equation. 

The approach above is Newtonian. More rigorous fully relativistic
derivations of the dispersion relation have been given by Ipser (1994,
1996), Perez et al.~(1997), and Silbergleit et al.~(2001).  Some
aspects of the general relativistic effects can be incorporated
into our analysis by using the Paczynski-Witta psuedo-Newtonian
potential, which gives $\kappa<\Omega=\Omega_\perp$.
For our purposes of estimating the c-mode damping rates,
it suffices to employ equation (\ref{mastereq}) 
but with the relevant fully general relativistic frequencies 
(e.g., Aliev \& Gal'tsov 1981; Okazaki et al.~1987)
\ba
\Omega &=& \frac{1}{r^{3/2} + a},\label{Omega}\\
\Omega_\perp &=& \Omega\left( 1 - \frac{4a}{r^{3/2}}+ \frac{3a^2}{r^2}\right)^{1/2},\\
\kappa &=& \Omega\left(1 - \frac{6}{r} + \frac{8a}{r^{3/2}} -
\frac{3a^2}{r^2}\right)^{1/2},\label{kappa},
\ea
where the frequencies are in units of $c^3/GM$, $r$ in units of $GM/c^2$, and
$a$ is the spin parameter of the black hole.

\section{Propagation Diagram and C-Modes}

\begin{figure}
\centering
\includegraphics[width=13cm]{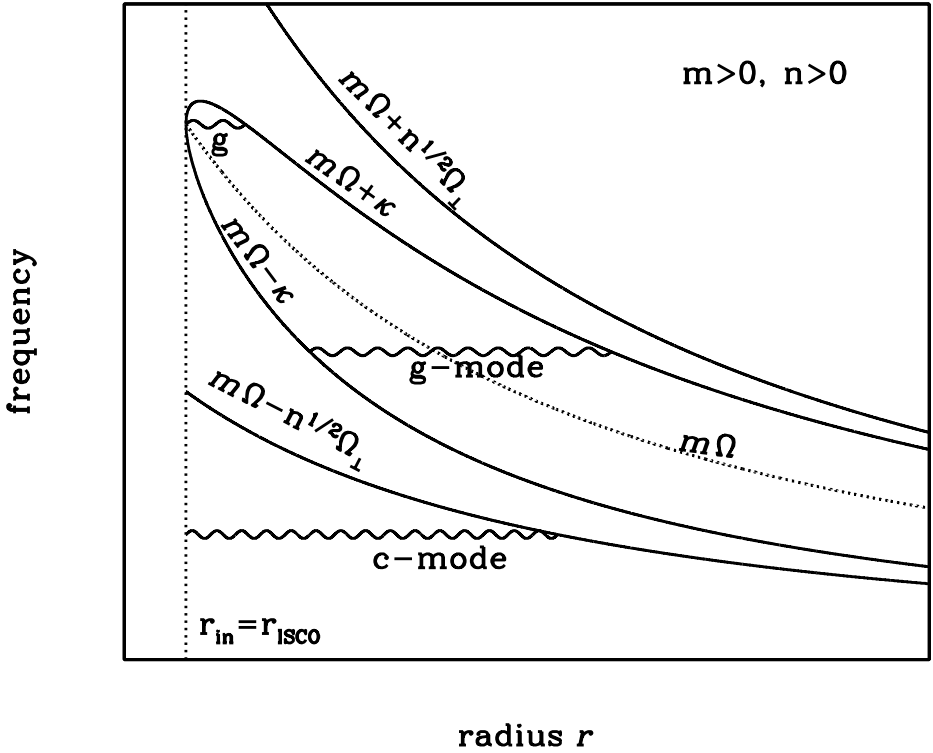}
\vskip -2.5cm
\caption{The propagation diagram for non-axisymmetric g-modes and
c-modes. Note that the curves $m\Omega-\kappa$ and 
$m\Omega + \kappa$ meet at $r=r_{\rm ISCO}$ since
$\kappa(r_{\rm ISCO}) = 0$ due to general relativistic effects.
The g-modes are trapped between the inner Lindblad resonance
(where $\omega=m\Omega-\kappa$) and the outer Lindblad resonance
(where $\omega=m\Omega+\kappa$), or around the peak of the $m\Omega+\kappa$
curve.}
\label{fig1}
\end{figure}
 
There are three possible critical resonant points in the disc:
the Lindblad resonances (LRs) where $D = 0$, the vertical resonances (VR) where
$\tomega^2 = n\Omega_\perp^2$, and the corotation resonance (CR) where
$\tomega = 0$. Far from these critical points, the WKB dispersion relation 
[for $\delta h\propto \exp(i\int k\, dr)$] takes the form (Okazaki et al.~1987)
\be
c_s^2k^2 = \frac{(\kappa^2 - \tomega^2)(n\Omega_\perp^2 -
\tomega^2)}{\tomega^2}.
\label{dispersion} \ee 
The modes with $n=0$ have no vertical structure, and are referred to
as p-modes -- their stability properties are studied in Lai \& Tsang
(2008).  We focus on the modes with $n\ge 1$ and $m\ge 1$ in this
paper.  Wave propagation is allowed in regions where
$\tomega^2<\kappa^2<n\Omega_\perp^2$ or
$\tomega^2>n\Omega_\perp^2>\kappa^2$ (note that $\kappa<\Omega_\perp$
in GR). The former defines the g-mode propagation zone: the mode is trapped
in the region where $m\Omega-\kappa<\omega<m\Omega+\kappa$; the latter leads to 
c-modes, for which the wave zone is specified by
$\omega < m\Omega - \sqrt{n}\Omega_\perp$ (see Fig.~1). Clearly, the c-modes
exist only when $m\Omega - \sqrt{n}\Omega_\perp>0$ and wave reflection
occurs at $r=r_{\rm in}=r_{\rm ISCO}$, 
the radius of the Inner-most Stable Circular Orbit.
For Newtonian discs, since $\Omega_\perp = \kappa = \Omega$ the
c-modes can only exist if $m > \sqrt{n}$; for relativistic
discs, however, $\Omega>\Omega_\perp$ (for spinning black holes), we can obtain
modes for $m = \sqrt{n}$ that may have very low frequencies.
The ``fundamental'' c-mode, with $m=n=1$ is of particular interest (Kato 1990), 
since it corresponds to the Lense-Thirring precession of the inner, tilted disc.
Note that for extremely low mode frequencies, the terms of order $1/r$
previously ignored may become important, and care must be taken to
obtain the real eigenfrequencies for trapped modes. Here we ignore
these complications and refer to Silbergleit et al.~(2001) for a more
thorough relativistic analysis.

\section{Wave Absorption at the Corotation Resonance}

The Lindblad resonances ($D=0$) are apparent singularities of the
master equation (\ref{mastereq}) [this can be seen easily by writing
(\ref{mastereq}) as two coupled first-order differential equations],
and no wave absorption occurs at the Lindblad
resonances (e.g., Goldreich \& Tremaine 1979; Li et al.~2003; 
Zhang \& Lai 2006). The vertical resonances (where $\tomega^2=n\Omega_\perp^2$)
act purely as turning points, and no wave absorption occurs there either. 
However, the corotation resonance must be treated more carefully
(Kato 2003; Li et al.~2003; Zhang \& Lai 2006). 

Here we follow the analysis of Zhang \& Lai (2006).
Near the corotation ($r=r_c$, where $\omega=m\Omega$),
equation (\ref{mastereq}) can be written as
\be
\frac{d^2}{dr^2}\delta h -\frac{D(\tomega^2 - n\Omega_\perp^2)}{c_s^2 \tomega^2} \delta h \simeq 0,\label{nearcr}
\ee
since for a thin disc the sound speed $c_s \ll r\Omega$ and
the last term in (\ref{mastereq}) dominates the other terms.
Defining $x\equiv (r-r_c)/r_c$ and expanding 
(\ref{nearcr}) around $x=0$, we have
\be
\frac{d^2}{dx^2}\delta h + \frac{C}{(x+i\epsilon)^2}\delta h = 0,\label{nearcrC}
\ee
where
\be
C \equiv \frac{n}{m^2}\left(\frac{\kappa\Omega_\perp}
{c_s d\Omega/dr}\right)^2_{r_c} \gg 1.
\ee
In equation (\ref{nearcrC}), we have inserted a small imaginary part $i\epsilon$
(with $\epsilon>0$) in $1/x^2$ because we consider the response of the
disc to a slowly growing perturbation.

The two independent solutions to equation (\ref{nearcrC}) are
\be
\delta h_\pm=z^{1/2}z^{\pm i\nu}
=z^{1/2}e^{\pm i\nu \ln z}
\label{c10}\ee
where $\nu=\sqrt{C-{1\over4}}\gg 1$ and $z=x+i\epsilon$.
The solution $z^{1/2}z^{i\nu}$ has a local wavenumber
$k=d(\nu\ln z)/dr=\nu/(r_c x)$, with the group velocity
$v_g=d\omega/dk=-\tomega/k=-q r_c x^2\omega/\nu<0$
(where we have assumed $\Omega\propto r^{-q}$, with $q>0$),
thus it represents waves propagating toward small $r$. Similarly,
the solution $z^{1/2}z^{-i\nu}$ has $v_g>0$ and represents waves propagating
toward large $r$.

As shown in Zhang \& Lai (2006), waves with $n\ge 1$ can propagate into
the corotation region and be absorbed there. Consider an incident
wave propagating from the $x<0$ (or $r<r_c$) region toward $x=0$, with 
the amplitude (up to a constant factor)
\be
\delta h(x<0)= z^{1/2}e^{-i\nu\ln z}=i e^{\pi\nu}(-x)^{1/2}e^{-i\nu\ln(-x)},
\qquad ({\rm incident~wave}).
\label{eq:incident2}\ee
The transmitted wave is simply
\be
\delta h(x>0)= x^{1/2}e^{-i\nu\ln x},\qquad
({\rm transmitted~wave}),
\ee
and there is no reflection.
Thus the amplitude of the transmitted wave is decreased by a factor 
of $e^{-\pi \nu}$ (Zhang \& Lai 2006).
Similarly, a wave incident from the $r>r_c$ region toward $r_c$ 
encounters the same attenuation. Since $\nu \gg 1$ for thin discs, 
We readily conclude that all waves incident upon the corotation will be 
absorbed (Kato 2003; Li et al.~2003; see Zhang \& Lai 2006 and Lai \& Zhang 2008
for applications of this result to the problem of wave excitation by 
external forces).

\section{Corotational Damping of C-modes}

\begin{figure}
\centering
\includegraphics[width=13cm]{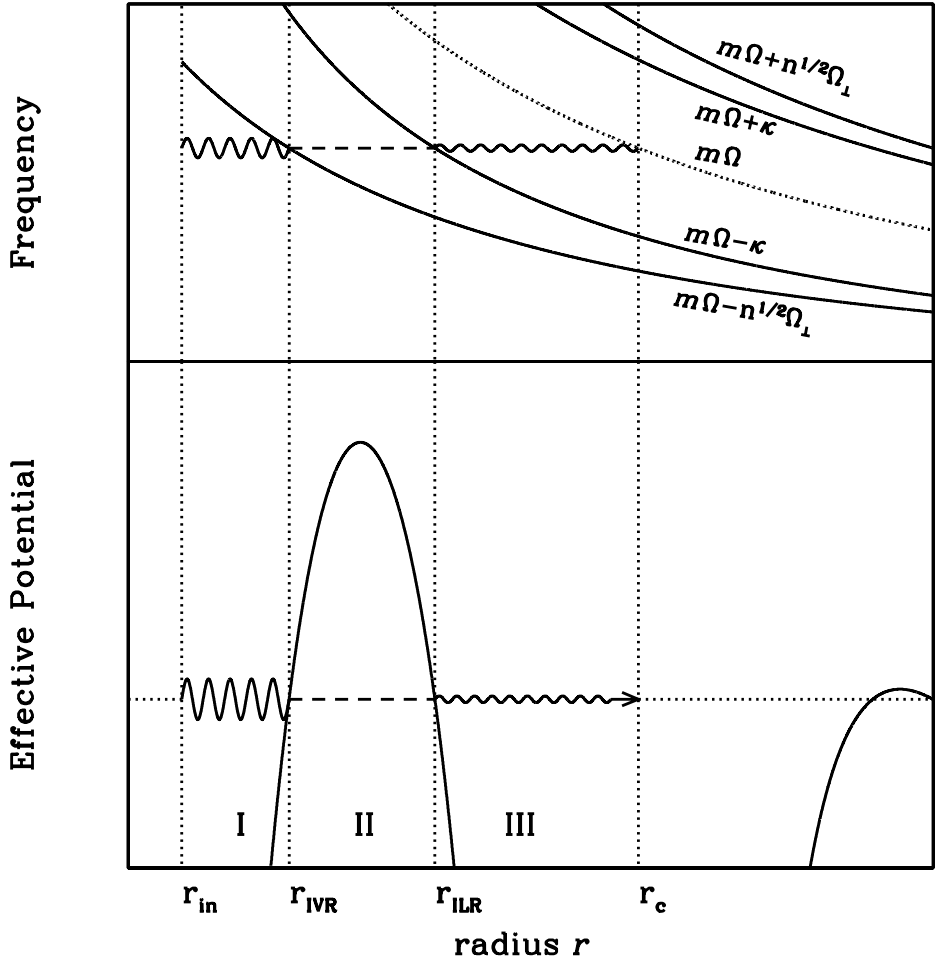}
\caption{The wave propagation diagram (upper panel) and effective potential 
(lower panel) for c-modes. The modes are trapped in region I (between $r_{\rm in}=
r_{\rm ISCO}$ and the inner vertical resonance $r_{\rm IVR}$), but can tunnel
through an evanescent barrier (region II, between 
the IVR and the inner Lindblad resonance $r_{\rm ILR}$)
and propagate into the corotation ($r_c$), where they are absorbed.}
\label{fig2}
\end{figure}

The result of section 4 shows that waves propagating through the
corotation are heavily damped. Since g-modes trapped between the inner
and outer Lindblad resonances must cross the corotation, they are
damped very quickly by corotation absorption, as shown by Kato (2003) and
Li et al.~(2003). Only higher frequency g-modes which
are trapped around the peak of $m\Omega+\kappa$ (see Fig.~1)
can avoid such corotational damping; these modes have frequencies
$\omega\simeq m\Omega(r_{\rm ISCO})$.

In this section we calculate the damping rate of the c-mode. The damping 
mechanism is illustrated in Fig.~2, where we also plot the effective potential:
\be
V_{\rm eff}(r)=  \frac{D(\tomega^2 - n\Omega_\perp^2)}{c_s^2 \tomega^2} + \frac{m^2}{r^2} +\frac{2m\Omega}{r\tomega}\left(\frac{d}{dr}\ln\frac{\Omega \Sigma}{D}\right).
\ee
Based on the result of section 4, we will adopt the approximation that
waves transmitted through the barrier between the inner vertical
resonance (IVR) and the inner Lindblad resonance (ILR) are absorbed at
the corotation. We will also assume that the inner disc boundary is
completely reflective (see section 6).

\subsection{Reflection Coefficient}

We first calculate the reflection coefficient ${\cal R}$ when a wave
in region I (see Fig. 2) propagates outward and is reflected back at the IVR.

From the dispersion relation (\ref{dispersion}), the group velocity
of the wave is given by 
\be
v_g\equiv {d\omega\over
dk}={kc_s^2\over \tomega \Bigl[
1-(\kappa/\tomega)^2(n\Omega_\perp^2/\tomega^2)\Bigr]}.
\label{gv}\ee
The relative sign of $v_g$ and the phase velocity $v_p=\omega/k$
is important. In region III ($r_{\rm ILR} < r < r_{c}$), $v_g$ and $v_p$ 
have the same sign, thus waves propagating outwards
correspond to $k > 0$. In region I ($r < r_{\rm IVR}$), 
$v_g$ and $v_p$ have opposite signs, so that 
waves propagating outwards have $k < 0$ and waves propagating inwards
have $k > 0$.

As shown in the previous section we can, to good approximation, assume
that waves propagating into the corotation are completely damped.
Thus, only an outward-propagating wave exists in region III (see Fig.~2), with
the wave amplitude (up to a constant prefactor) given by 
\be
\delta h =A\exp\left( i \int_{r_{\rm ILR}}^r k\,dr
+ \frac{\pi}{4}\right),
\label{eq:outside}\ee
where $k>0$ is given by equation (\ref{dispersion}) and
and $A \equiv \sqrt{D/r k \Sigma}$ is the WKB amplitude.
The connection formulae for the ILR (Tsang \& Lai 2008) are
\be
\delta h_1 \sim
\Biggl\{\begin{array}{ll}
\frac{1}{2}A\, \exp\left(-\int_r^{r_{\rm ILR}}\!|k| \,dr \right) 
& \qquad \textrm{for } r \ll r_{\rm ILR}\\
A\, \cos\left(\int_{r_{\rm ILR}}^r\! k\, dr + \pi/4\right) 
&\qquad \textrm{for } r\gg r_{\rm ILR}
\end{array}
\label{eq:con1}\ee
\be
\delta h_2  \sim
\Biggl\{\begin{array}{ll}
A \exp\left(\int_r^{r_{\rm ILR}}\! |k|\, dr \right) 
& \qquad \textrm{for } r \ll r_{\rm ILR}\\
A\, \sin\left(\int_{r_{\rm ILR}}^r\! k\, dr + \pi/4\right) 
&\qquad \textrm{for } r\gg r_{\rm ILR}~.
\end{array}
\label{eq:con2}\ee
These then give for the evanescent zone (region II in Fig.~2):
\ba
\delta h &\simeq&  \frac{A}{2}\exp\left(-\int_r^{r_{\rm ILR}}|k|\,dr 
\right) + iA\exp\left(\int_r^{r_{\rm ILR}} |k| dr\right) \nonumber\\
&=& \frac{A}{2}
\exp(-\Theta_{\rm{II}}) 
\exp\left(\int_{r_{\rm IVR}}^{r}|k| dr \right) + iA
\exp(+\Theta_{\rm{II}})\exp\left(-\int_{r_{\rm IVR}}^r |k| dr\right) 
\ea
where 
\be
\Theta_{\rm{II}} = \int_{r_{\rm IVR}}^{r_{\rm ILR}} |k|\, dr.
\ee

\begin{figure}
\centering
\includegraphics[width=17cm]{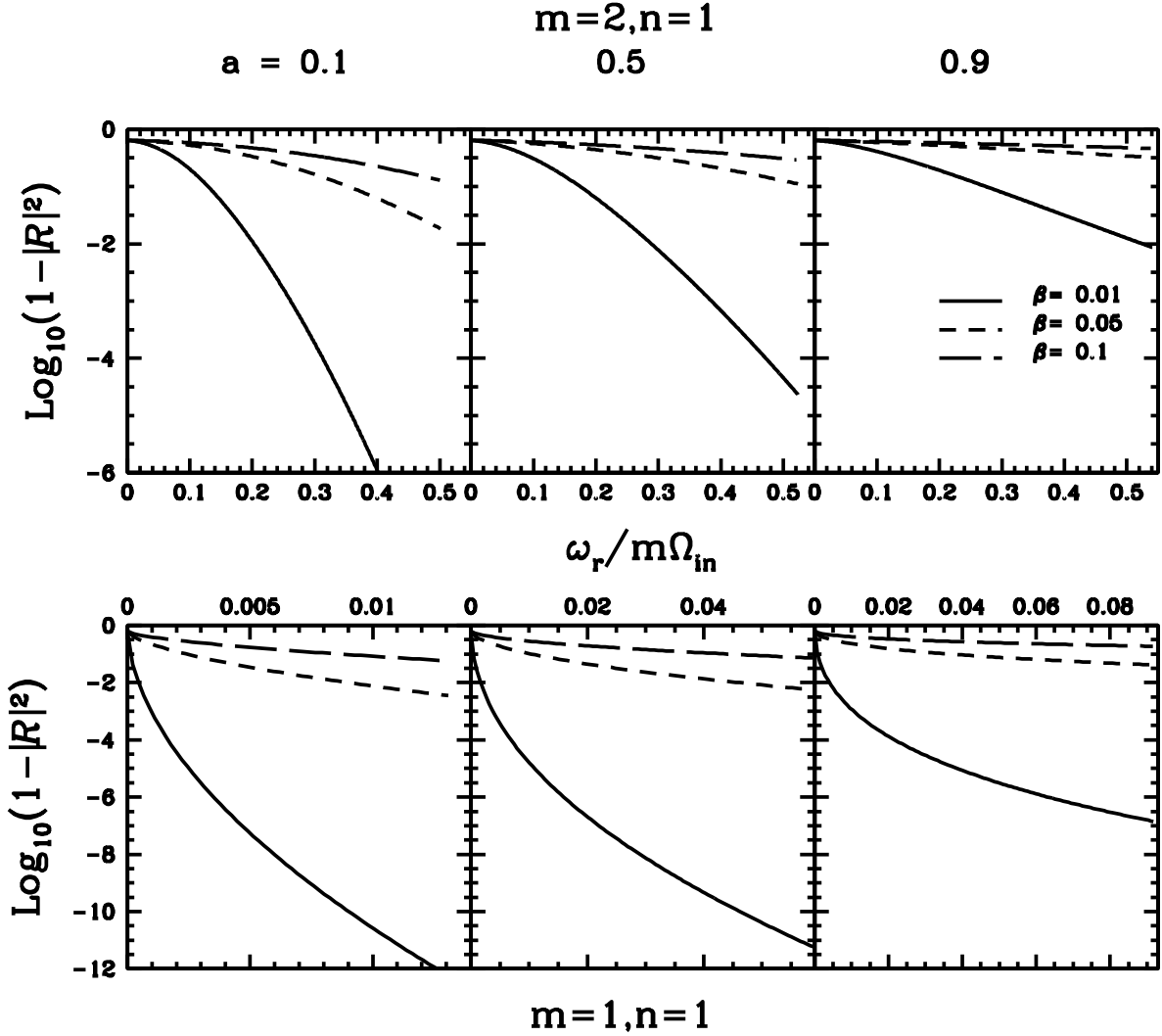}
\vskip -2cm
\caption{The reflection coefficient, ${\cal R}$, of waves incident
upon the inner vertical resonance (IVR), as a function of the (real)
wave frequency $\omega_r$ (in units of $m\Omega_{\rm in}$, where $\Omega_{\rm in}$
is the disc rotation rate at $r=r_{\rm in}$), for
various black hole spin parameters ($a$) and sound speeds $c_s = 
\beta r \Omega_{\rm in}$. The upper panels are for $m=2,~n=1$ and
the lower panels for $m=n=1$.}
\label{fig3}
\end{figure}

The connection formulae at the IVR can be similarly derived; they are
\footnote{Note that around the IVR, the differential equation is matched
by the Airy functions, rather than by the Airy function derivatives
as in the case of the ILR.}
\be
\delta h_1 \sim
\Biggl\{\begin{array}{ll}
\frac{1}{2}A\, \exp\left(-\int^r_{r_{\rm IVR}}\!|k| \,dr \right) 
& \qquad \textrm{for } r \gg r_{\rm IVR}\\
A\, \sin\left(\int^{r_{\rm IVR}}_r\! k\, dr + \pi/4\right) 
&\qquad \textrm{for } r\ll r_{\rm IVR}
\end{array}
\label{eq:conn1}\ee
\be
\delta h_2  \sim
\Biggl\{\begin{array}{ll}
A \exp\left(\int^r_{r_{\rm IVR}}\! |k|\, dr \right) 
& \qquad \textrm{for } r \gg r_{\rm IVR}\\
A\, \cos\left(\int^{r_{\rm IVR}}_r\! k\, dr + \pi/4\right) 
&\qquad \textrm{for } r\ll r_{\rm IVR}
\end{array}
\label{eq:conn2}\ee
(Abramowitz \& Stegun 1964). 
Thus, we find that for $r < r_{\rm IVR}$ (region I in Fig.~2),
\be
\delta h \simeq \frac{A}{2}e^{-\Theta_{\rm{II}}}
\cos\left(\int_r^{r_{\rm IVR}} k\,dr +\frac{\pi}{4}\right) + i2A
e^{\Theta_{\rm{II}}}\sin\left(\int_r^{r_{\rm IVR}} k\,dr 
+\frac{\pi}{4}\right)~.
\ee
Expressing this in terms of traveling waves and defining 
$y = \int_{r_{\rm IVR}}^r k dr - \pi/4$, we have
\be
\delta h \simeq iA \left[ e^{-iy} \left(e^{+\Theta_{\rm{II}}} 
+ \frac{1}{4}e^{-\Theta_{\rm{II}}}\right) - e^{+iy} \left(
e^{\Theta_{\rm{II}}} - \frac{1}{4}e^{-\Theta_{\rm{II}}}\right)\right],
\label{eq:inside}\ee
where the first term ($\propto e^{-iy}$) corresponds to the incident (outgoing) wave
and the second term ($\propto e^{iy}$) the inward going wave reflected from the IVR.
Thus the reflection coefficient is 
\be
{\cal R} = -\frac{e^{\Theta_{\rm{II}}} - \frac{1}{4}e^{-\Theta_{\rm{II}}}}{e^{\Theta_{\rm{II}}} +
\frac{1}{4}e^{-\Theta_{\rm{II}}}},
\ee
and the transmission coefficient (through region II) is
\be
{\cal T} = \frac{i}{e^{\Theta_{\rm{II}}} +
\frac{1}{4}e^{-\Theta_{\rm{II}}}}.
\ee
We can clearly see that $|{\cal R}|^2 < 1$.

\subsection{Trapped C-modes and their Damping Rates}

Assuming a reflective boundary exists at $r = r_{\rm in} < r_{\rm IVR}$,
we can develop trapped c-modes in the inner disc, between $r_{\rm in}$ and 
the IVR. To illustrate this we consider a simple boundary condition at $r=r_{\rm in}$:
\be
\delta h(r_{{\rm in}}) = 0. \label{boundary}
\ee
From section 5.1, the wave in region I can be written as 
\be
\delta h=A\exp(-iy)+{\cal R}A\exp(iy),\qquad {\rm with}~~
y=\int_{r_{\rm IVR}}^{r} \!k\,dr-\pi/4.
\label{eq:rinside} \ee
where $k=k_r+ik_i$ is complex.
Applying the boundary condition (\ref{boundary}) to 
equation (\ref{eq:rinside}) yields the eigenvalue condition:
\be 
\exp(2i\Theta)=-i|{\cal R}|,\quad {\rm with}~~
\Theta=\int_{r_{\rm in}}^{r_{\rm IVR}}\!k\,dr=\Theta_r+i\Theta_i,
\label{eq:eigcond}
\ee
where $\Theta_r$ and $\Theta_i$ are real.
The real eigen frequency $\omega_r$ is given by
\be
\Theta_r=\int_{r_{\rm in}}^{r_{\rm IVR}}\!k_r\,dr
=\int_{r_{\rm in}}^{r_{\rm IVR}} 
{\sqrt{(\kappa^2 - \tomega_r^2)(n\Omega_\perp^2 - \tomega_r^2)}\over 
c_s |\tomega_r|}\,dr
=\mu \pi +{3\pi\over 4},
\label{omegarcond}\ee
where $\mu=0,1,2,\cdots$ is an integer and 
$\tomega_r = \omega_r - m\Omega < 0$ in the trapping region
between $r_{\rm in}$ and $r_{\rm IVR}$.
The imaginary part of the frequency $\omega_i$ is determined by
$|{\cal R}|=\exp(-2\Theta_i)$, or
\be
\tanh\Theta_i=-\left({|{\cal R}|-1\over |{\cal R}|+1}\right).
\ee
Note that $\Theta_i=\int_{r_{\rm in}}^{r_{\rm IVR}}k_i\,dr\ll 1$ and
\be
k_i=\omega_i {dk\over d\omega}\Bigr|_{\omega_r}=
\frac{\omega_i\tomega_r}{k_r c_s^2}\left(1-
\frac{n\kappa^2\Omega_\perp^2}{\tomega_r^4} \right).
\ee
We then obtain
\be
\omega_i \simeq -\frac{1}{4}\exp(-2\Theta_{\rm II})
\left[\int_{r_{\rm in}}^{r_{\rm IVR}}\!\frac{|\tomega_r|^2}
{\sqrt{(\kappa^2 -\tomega^2)(n\Omega_\perp^2 - \tomega^2)}}
\left(1 -\frac{n\kappa^2\Omega_\perp^2}{\tomega_r^4}\right)
{dr\over c_s}\right]^{-1}.
\label{omegai} 
\ee
Note that in the trapping region (between $r_{\rm in}$ and $r_{\rm IVR}$),
$n\kappa^2\Omega_\perp^2/\tomega^4 < 1$. Thus the mode is always damped
($\omega_i<0$).

\begin{figure}
\centering
\begin{tabular}{ccc}
\epsfig{file=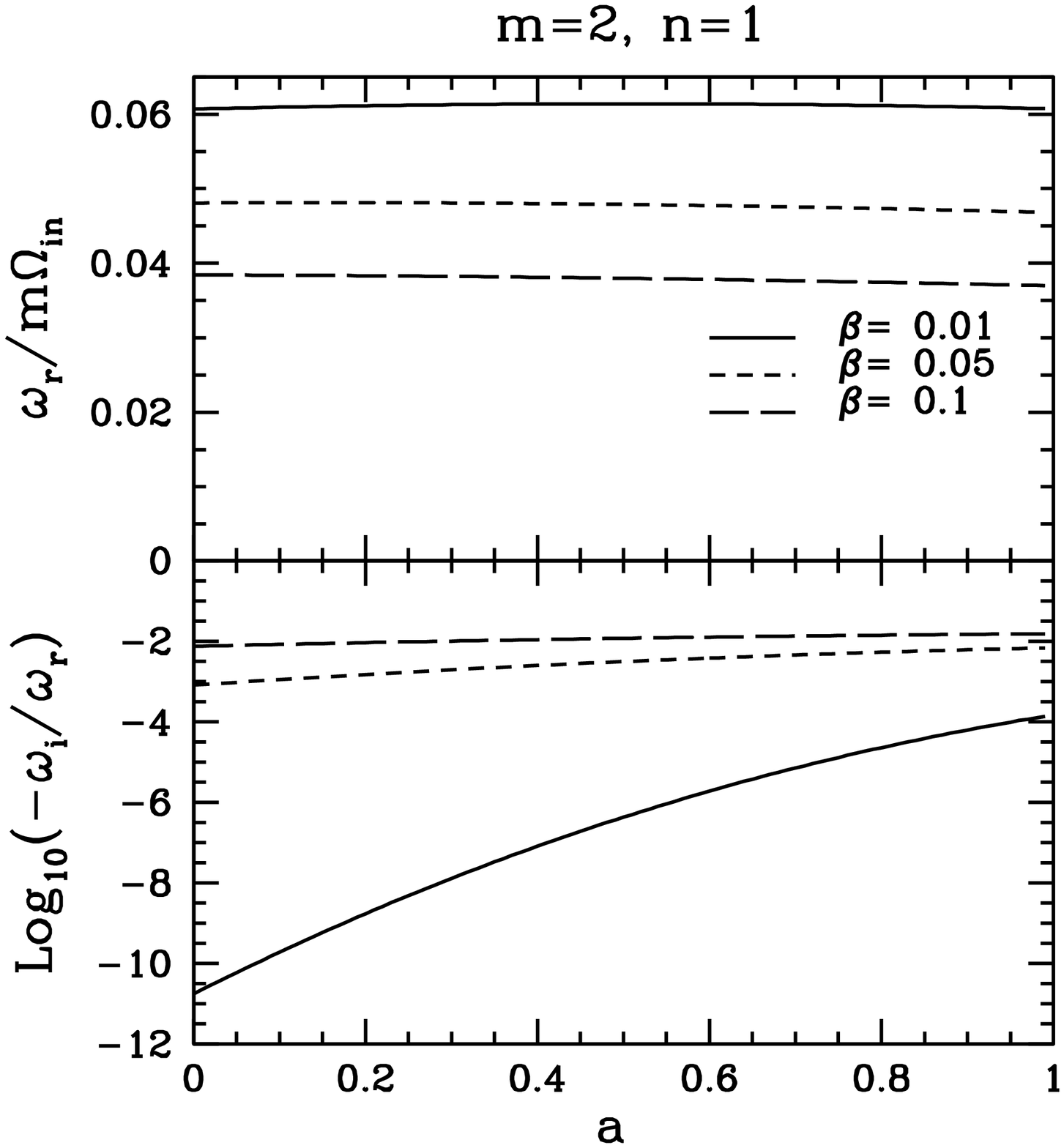,width=0.5\linewidth,clip=} &
\epsfig{file=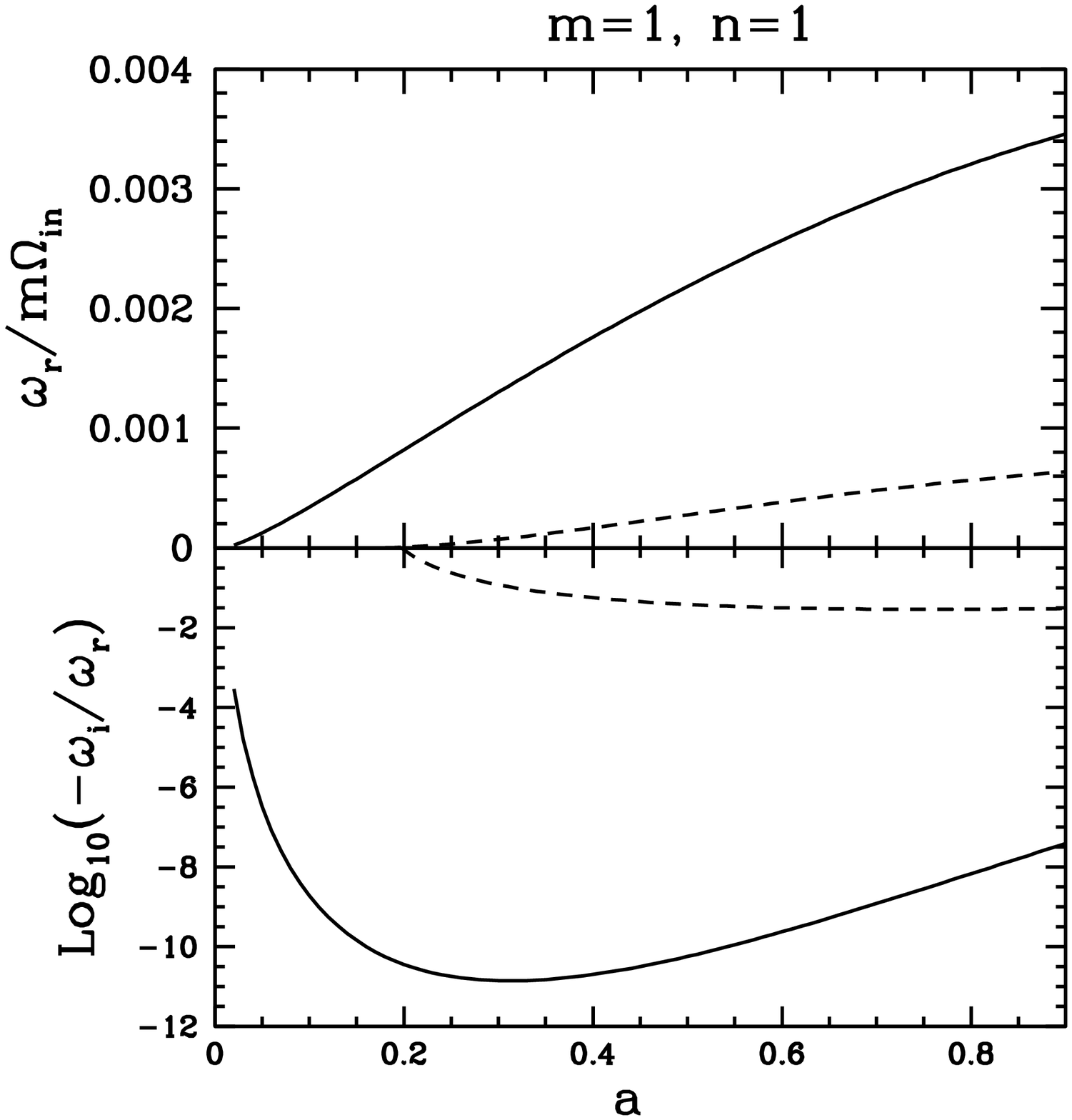,width=0.5\linewidth,clip=}
\end{tabular}
\caption{The real and imaginary frequencies for the primary trapped
c-mode with the $\delta h(r_{\rm in}) = 0$ boundary condition and
$r_{\rm in} = r_{\rm ISCO}$ for various sound speeds versus the black
hole spin parameter. Note that there is no mode for the one-armed
corrugation wave (the right panel) for sound speed $c_s = 0.1
r\Omega_{\rm in}$.}
\label{fig4}
\end{figure}

\subsection{Numerical Results}

Figure 3 depicts the reflection coefficients for waves impinging
upon the inner vertical resonance (IVR) from $r < r_{\rm IVR}$
as a function of the wave frequency. We consider both the 
$m=2,~n=1$ and the $m=1,~n=1$ modes. The real frequency ranges from
$0$ to $(m\Omega-\sqrt{n}\Omega_\perp)|_{r_{\rm in}}$. 
Waves with higher frequencies are ``protected'' by a larger potential barrier in the evanescent zone (larger $\Theta_{\rm II}$) and have $|{\cal R}|^2$ closer to unity.

Figures 4-5 show the real and imaginary c-mode frequencies, computed
using the WKB expressions derived in section 5.2. Waves with 
higher frequencies have smaller damping rate $|\omega_i|$, consistent with 
the reflectivity results shown in Fig.~3. The damping rate is also smaller for 
cooler (smaller $\beta$) discs. 
For the $m > n$ case (e.g. Fig. 4, left panel), the dimensionless damping rate $|\omega_i|/m\Omega_{\rm in}$ increases with increasing $a$ (while $\omega_r/m\Omega_{\rm in}$ remans approximately constant), because the width of the region between $r_{\rm IVR}$ and $r_{\rm ILR}$ is smaller for larger $a$. For the $m = n = 1$ case (Fig. 4, right panel) the range of possible c-mode frequencies is bounded from above by the Lense-Thirring precession frequency at the inner boundary. Thus $\omega_r \rightarrow 0$ when $a \rightarrow 0$. This leads to larger $|\omega_i/\omega_r|$ for small $a$ since the potential barrier in the evanescent zone decreases with decreasing $\omega_r/m\Omega_{\rm in}$.
Note that although we use the relativistic
frequencies (\ref{Omega})-(\ref{kappa}) our calculations not fully
relativistic, and the real frequencies calculated shown in Fig.~4, particularly for
the $m=n=1$ mode with large black hole spin parameter $a$, are 
correct only in orders of magnitude [Fully relativistic calculation of the real
frequencies of these trapped modes was done
by Silbergleit et al.~(2001).] Also note that adopting different inner disc boundary 
conditions would lead to different real mode eigenfrequencies than presented in
Fig.~4. 

Finally, we note that equation (\ref{omegai}) is valid only when there
is no loss of wave energy at the inner disc boundary $r_{\rm in}$.
In the presence of the rapid radial inflow at the ISCO, we would expect 
additional mode damping due to the leakage of waves into the plunging region
of the disc (see Lai \& Tsang 2008 for a study of such leakage for p-modes).
Alternatively, a sufficiently strong excitation mechanism is needed to 
overcome the corotational damping and make the c-modes grow.

\begin{figure}
\centering
\includegraphics[width=17cm]{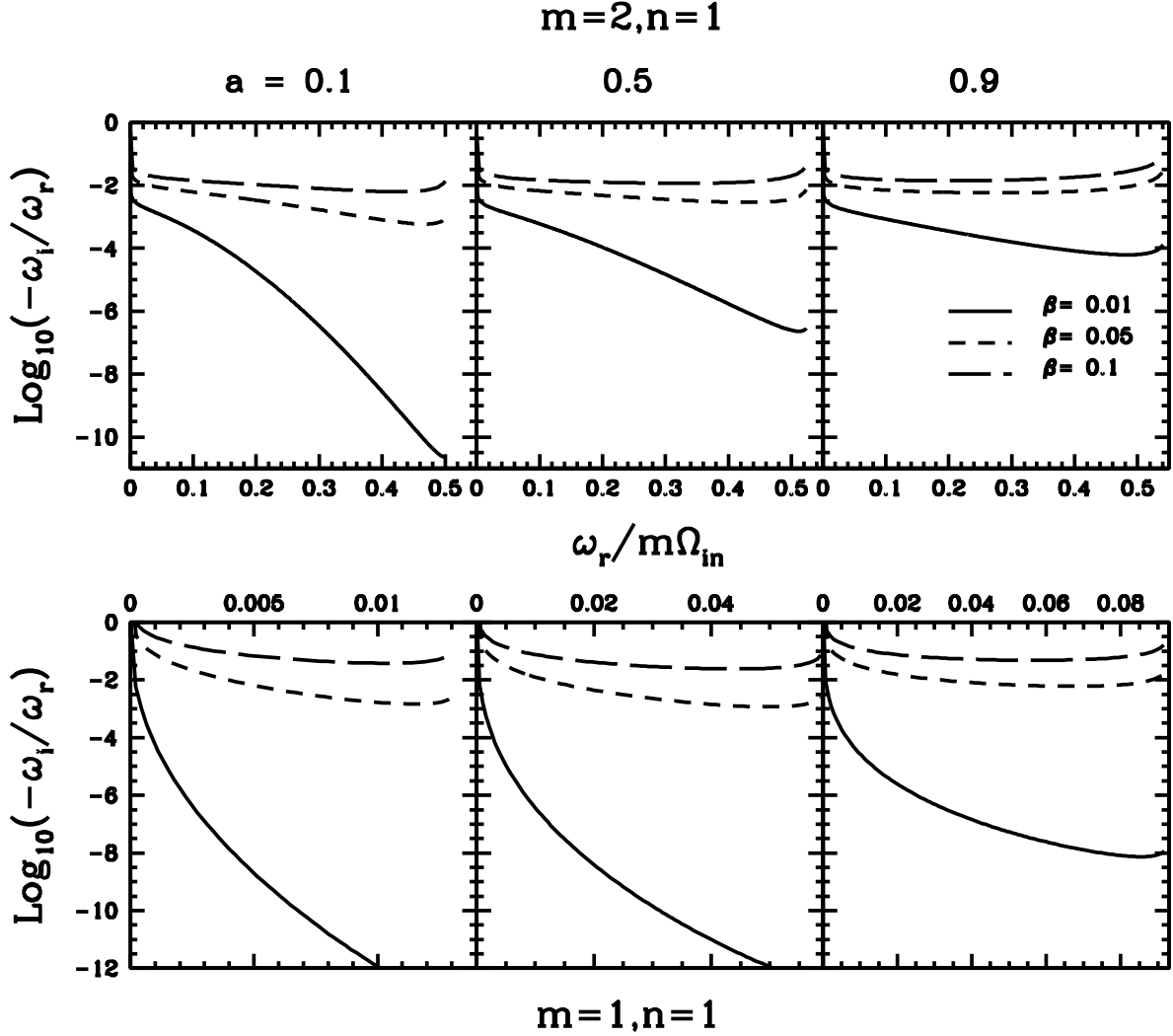}
\vskip -2cm
\caption{The damping rate $|\omega_i|$ for various sound speeds and
black hole spin parameters for c-modes with purely reflective boundary
conditions at $r_{in}$ as a function of the real mode frequencies $\omega_r$. 
The upper panels show the results for the $m=2,n=1$ modes and the lower panels
show the $m=n=1$ modes. The sound speeds are taken to be constant,
$c_s = \beta r\Omega_{\rm in}$.
}\label{fig5}
\end{figure}

\section{Conclusion}

In this paper, we have shown that diskoseismic c-modes suffer
corotational damping due to wave absorption at the corotation
resonance. These modes are trapped between the inner disc edge and the inner
vertical resonance (where $\omega-m\Omega=-\sqrt{n}\Omega_\perp$, 
with $m,n\ge 1$), but can tunnel through the evanescent zone and leak
out to the corotation zone where wave absorption occurs.
The mode damping rates are generally much smaller than the mode
frequencies, and depend sensitively on the disc sound speed and
the black hole spin parameters.

With this paper, we now have in hand a complete picture of how the 
corotation resonance affects various diskoseismic modes 
in black-hole accretion discs, at least in the linear regime:
Non-axisymmetric g-modes are heavily damped at the corotation resonance
(Kato 2003; Li et al.~2003), while p-modes (inertial-acoustic modes) 
can be overstable due to the corotational wave absorption (see Lai \& Tsang
2008 and references therein). The corotational damping rates of c-modes
are much smaller than those of g-modes.

Diskoseismic c-modes have been invoked to explain low-frequency
variabilities in black-hole X-ray binaries (van der Klis 2006;
Remillard \& McClintock 2006). Fu \& Lai (2008) showed that the basic
properties of c-modes are largely unaffected by the disc magnetic
fields and thus these modes are present in real discs. The results
presented in this paper show that in order for the c-modes to be
observable, a sufficiently strong excitation mechanism is needed to
overcome the corotational damping.

\section*{Acknowledgments}

This work has been supported in part by NASA Grant NNX07AG81G, NSF
grants AST 0707628, and by {\it Chandra} grant TM6-7004X
(Smithsonian Astrophysical Observatory).


\end{document}